\documentstyle[11pt,newpasp,twoside,epsf]{article}
\def\edcomment#1{\iffalse\marginpar{\raggedright\sl#1\/}\else\relax\fi}
\reversemarginpar

\begin{document}
\title{Heating of Gas in Galaxy Groups and
Clusters}
\author{Fabrizio Brighenti}
\affil{Dipartimento di Astronomia, Universit\`a di Bologna, via Ranzani 1,
Bologna 40127, Italy, brighenti@bo.astro.it}
\author{William G. Mathews}
\affil{University of California Observatories/Lick Observatory,
Board of Studies in Astronomy and Astrophysics, University of California,
Santa Cruz, CA 95064, mathews@ucolick.org}

\begin{abstract}
Was the diffuse gas in galaxy groups and clusters heated at high
redshift before
it entered a massive halo, or was the heating produced inside 
collapsed objects
by SNII accompanying normal star formation?
We compare here two radically different models corresponding to the
two scenarios described above. Our results indicate that internal heating
by SNII works better than the extreme version of external heating
that we adopt in reproducing the observed $L-T$ and ${\rm entropy}-T$
relations.
\end{abstract}

\section{Introduction}

In a simple universe governed only by gravity, the 
bremsstrahlung luminosity of galaxy
clusters would scale as $L\propto T^2$, where $T$ is the gas temperature
(Kaiser 1986). However, the observed relation is steeper, $L\propto T^3$
or $L\propto T^4$ (e.g. Arnoud \& Evrard 1999; Helsdon \& Ponman 2000).
Other relations, notably the one between entropy and temperature, also show
deviation from self-similarity (Lloyd-Davies, Ponman, \& Cannon 2000;
Mohr, Mathiesen, \& Evrard 1999).

These discrepancies can be explained if gas in clusters
experienced some non-gravitational
heating before (or when) it flowed into the cluster halo
(Kaiser 1991; Evrard \& Henry 1991). However, there is no general consensus
about the details of the heating process.

Here we compare two pictures for the heating that have been proposed
by a number of authors in recent years: the {\it external heating} scenario,
in which the gas is heated at high redshifts, before it enters
massive halos, and the {\it internal heating} scenario, in which
the energy is injected when some or most of
the gas is already inside a massive halo.
In this latter case we assume that the sources of energy are SNII
explosions. 
The motivation for considering SNII heating a viable
form of heating is the success of our models
in reproducing the evolution of ISM
in NGC 4472 (Brighenti \& Mathews 1999a).
NGC 4472 is a giant elliptical at the center of a
subcluster in Virgo, consistent with being the remnant
of a galaxy group, probably stripped of the outer regions when it entered
the Virgo cluster. We showed that SNII heating was able to reproduce
the gas density and temperature profiles 
(and therefore also the entropy
profile).

\section{The Simulations}
We use a modified version of the hydrocode ZEUS (Stone \& Norman 1992)
and we assume spherical symmetry. The code follows two fluids: a normal,
collisional gas, and a collisionless fluid which represents the dark matter.
Our groups and clusters evolve from a single top-hat perturbation
in a $\Lambda$CDM universe ($\Omega_0 = 0.3$, $\Lambda = 0.7$,
$h = H_0/(100~{\rm km/s~Mpc}) = 0.725$, $\Omega_{\rm b}=0.037$).
The dark matter accumulates in a Navarro, Frenk, \& White (1996)
halo by design. We consider a set of three objects with different virial 
masses (at the present time):

- $M_{\rm vir} = 4.7 \times 10^{13}$ M$_{\sun}$: ``the group''; \par
- $M_{\rm vir} = 2.2 \times 10^{14}$ M$_{\sun}$: ``the poor cluster''; \par
- $M_{\rm vir} = 1.2 \times 10^{15}$ M$_{\sun}$: ``the rich cluster''; \par

\noindent We focus here mainly on groups, since lower mass system
are more sensitive to heating, and our models, by ignoring
the complex merging events,
are anyway less appropriate to describe the formation and evolution
of large clusters.

{\it \underline{External pre-heating}}.
We assume an extreme form of preheating:
at very high redshift, $z_{\rm h}=9$, we reset everywhere 
the gas density to the mean
baryon density, $\rho = {\bar \rho}_{\rm b}(z=9)$. 
At that epoch, the temperature
is raised to some constant level $T_{\rm h}$. 
We consider 4 levels of heating: $T_{\rm h} = 10^4, 5\times 10^6, 
10^7, 3\times 10^7$ K, corresponding to $1.3 \times 10^{-4}, 0.65,
1.3, 3.9$ keV/particle. These amount of heating is
characterized by the numbers $1$ to $4$ respectively,
in our nomenclature.
The entropy parameter $S=T/n_{\rm e}^{2/3}$ corresponding to these
levels of heating, which depends on the heating epoch $z_{\rm h}$
through $n_{\rm e}$, is $S=0.025, 125, 250, 750$ keV cm$^{2}$.

Gas is allowed to cool and to dropout of the flow. This last process
is modeled in the usual way (e.g. Sarazin \& Ashe 1989) adding a sink
term in the continuity equation 
${\dot \rho}_{do} = -q \rho/t_{\rm cool}$ with
$q=1$.

{\it \underline{Internal heating}}. 
In this series of models, heating is assumed 
to be the result of star formation occurring inside the group
or cluster. Thus, we need to assume a schematic scenario for
star formation in these systems. At $z_* = 3$ (2 Gyr after the big-bang)
we form stars from cooled gas (conserving baryons) and release SNII
energy inside the accretion shock radius $r_{\rm sh}(z_*)$.
All the gas inside $r_{\rm sh}$ is assumed to get the same amount of
energy per unit mass. Te total amount of energy released
is $E_{\rm SN} = \nu \eta _{\rm Salpeter} E_0 (M_*/M_{\sun})$,
where $\eta _{\rm Salpeter}\sim 0.007$ is the number 
of SNII per unit solar mass
predicted by a stellar population with a Salpeter IMF, $E_0=10^{51}$ erg
is the kinetic energy released by a single SNII and $M_*$ is the total
mass of stars. The parameter $\nu$ controls the amount of heating
(i.e. the number of SNII).
We consider 4 models with $\nu = 0.5, 1, 2, 4$,
labeled with numbers 1 to 4, in analogy with the external heating models.

In more familiar units, SNII inject $\sim 2.4 \nu$ keV/particle
to the gas inside $r_{\rm sh}$ at $z_{\rm h}$.
This is consistent with 
the global value $0.22$ keV/particle which is derived
assuming a global star formation efficiency $M_*/M_{\rm baryon}=0.1$
(Fukugita, Hogan, \&
Peebles 1998) and assuming that the SNII energy is shared among all
the baryons.
In our models, instead, SNII heat only the central part of clusters,
similar to the models proposed by Loewenstein (2000).
In these models the hydrodynamic equations are modified to take into
account the mass and energy injected by stars and SNIa of the central,
dominant galaxy.
Full details about the simulations can be found in Brighenti \& Mathews (2001).

\begin{figure}
\vspace{4.5 truecm}
\includegraphics{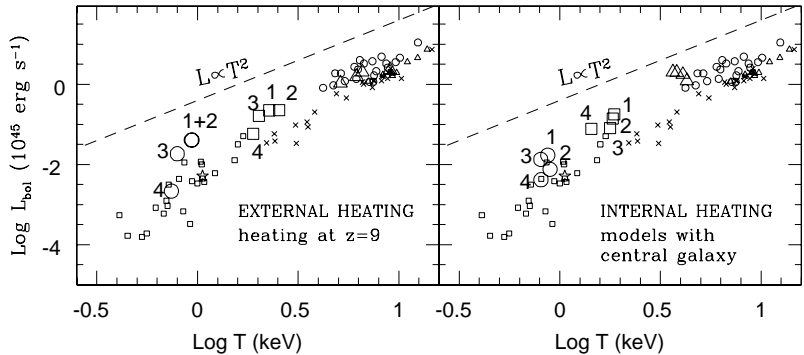}
\vspace{0.7 truecm}
\caption{Bolometric X-ray 
luminosity vs. emission weighted temperature for
{\it external heating} (left) and 
{\it internal heating} (right) models. Small symbols are data: open squares, 
crosses and open circles
are from Helsdon \& Ponman (2000), Arnoud \& Evrard (1999) and
Allen \& Fabian (1998) respectively. Big circles, squares and triangles
are models for groups, poor clusters and rich clusters. Labels 
$1\rightarrow4$ indicate the heating level (see text).}
\label{fig-1}
\end{figure}

\section{Results}

\subsection{The $L-T$ relation}

After evolving to the current time, $t = 13$ Gyrs,
the location of our models can be compared with 
observed clusters in the $L-T$ plot 
as shown in Fig. 1. In general, heating has a little effect on
the emission weighted temperature (heated clusters are not hotter!).
Paradoxically, models with maximum heating are always the coolest.
Instead, the luminosity generally decreases as heating increases, a result
of the lower mean gas density. Groups, having lower virial temperatures,
are more affected by the heating, while rich clusters are quite
insensitive to it.

The left panel of Fig. 1 shows the results for the {\it external heating}
scenario. Models ``1'' without heating nicely follow the self-similar
prediction $L\propto T^2$. 
The group model with maximum heating (model ``4'')
lies among the observed
groups, and the series of models ``4'' follows a relation $L\propto T^3$,
similar to the observed one. However, it requires that $\sim 3.9$
keV/particle are dumped in the gas, or that a entropy floor
$S\sim 750$ keV cm$^2$ is established at $z_{\rm h}=9$. The energy budget
needed is a function of $z_{\rm h}$. If $z_{\rm h}=5$, $\sim 1$ keV/particle
is sufficient to decrease the luminosity of groups to
the observed level.

The {\it internal heating} models are less sensitive to the amount
of energy injected (right panel of Fig. 1) and they tend to lie near the
upper envelope of the observed $L-T$ data, where strong cooling flows
are dominated by a massive central galaxy as we have assumed. 
Groups models with $\nu\ge 1$ have X-ray properties consistent with
observed groups, provided the efficiency of SNII heating is high (we 
assumed efficiency $=1$).
However, it is likely that a significant fraction of the SNII energy
is lost by radiation, and a more realistic constraint may be $\nu\ge 2-3$.

\subsection{The $S-T$ relation}

The behavior of entropy for {\it external heating} models is
illustrated in Fig. 2. In the left panel we compare the entropy
evaluated at $r=0.1 r_{\rm vir}$ with the data of Lloyd-Davies et al.
(2000). The series with maximum heating (models ``4'',
indicated with the largest circles), which fits best in the $L-T$ plot,
has a much larger entropy than observed groups and poor clusters.
Models ``3'' (which requires $\sim 1.3$ keV/particle) may be the
best compromise.

The difficulty for external heating models to simultaneously fit
the $L-T$ and $S-T$ relations is also illustrated by the radial entropy
profiles (right panel of Fig. 2). 
The computed group entropy profiles are compared
with the observed profile of NGC 2563 group (dot-dashed line; 
Trinchieri, Fabbiano,
\& Kim 1997); see also the profiles in Lloyd-Davies et al. (2000).
Real groups have entropy profiles 
that increase monotonically with radius,
while for model ``4'' $S$ is uniform and model ``3'' shows
a flat entropy core for $r<300$ kpc. This behavior is due to
the inefficiency of radiative cooling in the low density cores
of these strongly heated groups.

{\it Internal heating} group models fall nicely among the observations
in the $S-T$ plane (Fig. 3), regardless of the heating parameter $\nu$.
Strong radiative cooling regulates the entropy in the central regions
to have similar values for all adopted $\nu$.
Radial entropy profiles are also consistent with observations (Fig. 3,
bottom panel), although our models are somewhat denser overall with
slightly lower $S(r)$.

\begin{figure}
\vspace{4.5 truecm}
\includegraphics{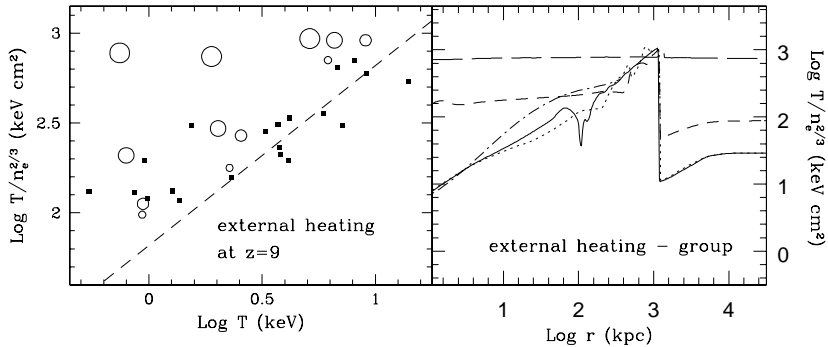}
\vspace{0.7truecm}
\caption{External heating. {\it Left:}
$S=Tn^{-2/3}$ evaluated at $0.1 r_{\rm vir}$
versus emission weighted temperature (open circles). The size of the
symbols increases with the amount of heating.
Data (filled squares) are taken
from Lloyd-Davies et al. (2000). {\it Right:} $S(r)$ profiles for group
models. Models 1$\rightarrow$4 are represented with solid, dotted,
short-dashed and long-dashed lines respectively. The dot-dashed line
is the observed profile for group NGC 2563.}
\label{fig-1}
\end{figure}

\subsection{Baryon fractions and cooling times}

Models experiencing external 
preheating at very early times differ from
those heated internally by SNII in several other respects.
Notably, it appears that internal heating removes baryons more
efficiently than preheating at $z_{\rm h}=9$.
We find for preheated groups that the baryon fraction at 
$r_{\rm vir}\sim 900$ kpc is $f_{\rm b}\sim 0.11$ for model ``3'', almost
equal to the cosmic baryon fraction assumed, $0.123$. Only for the
maximally preheated model ``4'', is the baryon fraction 
significantly lower than the cosmic one: $f_{\rm b}\sim 0.07$.

Groups heated internally by SNII have low baryon fraction even
when a Salpeter IMF is assumed ($\nu=1$):
$f_{\rm b}\sim 0.075$ ($f_{\rm b}\sim 0.055$ for $\nu=2$).
The fraction of mass in gas
at $r_{\rm vir}$ is $\sim 0.04$ and $\sim 0.03$ for 
$\nu=1$ and $\nu=2$, respectively.

A further distinction between external and internal heating models
is the central cooling time which, if lower than the age of the
system, may indicate the presence of a cooling flow.
We find that models with preheating strong enough to fit
observations in the $L-T$ plot (models 3 and 4 in Fig. 1) 
never develop cooling flows (or, more precisely, never have
$t_{\rm cool}<~{\rm age}$), contrary to many observed groups and clusters.
All internal heating models, instead, develop strong cooling
flows. In particular, group models have $\dot M \approx 1-10$ 
M$_{\sun}$ yr$^{-1}$.

\begin{figure}
\vspace{4.5truecm}
\includegraphics{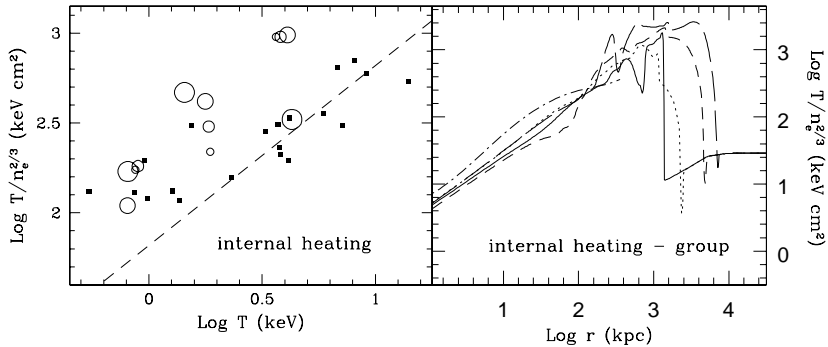}
\vspace{0.7 truecm}
\caption{Same as Fig. 2 for internal heating models.}
\label{fig-1}
\end{figure}

\section{Heating-enrichment connection}
Both the external and internal heating scenarios seem to require more energy
than that provided, via SNII, by the observed stellar content
with a Salpeter IMF. A reasonable requirement may be $\sim 1$ keV/particle
(a value consistent with most models proposed in recent years).
SNII also produce metals, so we should ask:
is the number of SNII needed to heat the gas consistent
with the number of SNII necessary to produce the metal content of the
universe?

A Salpeter IMF (from 0.1 to 100 M$_{\sun}$) produces $\sim 0.007$ SNII
per M$_{\sun}$ of stars formed. Assuming a global star formation efficiency
$\Omega_*/\Omega_{\rm baryon}=0.1$ (Fukugita et al. 1998), we get
$\epsilon \sim 0.22 \nu$ keV/particle. Thus, to generate $\sim 1$ keV/particle
we need $\nu\sim 5$.

The present day universe has a global iron abundance $<Z_{\rm Fe}>
\sim 0.3-0.4$ solar meteoritic units (Renzini 1997).
Assuming again $\Omega_*/\Omega_{\rm baryon}=0.1$
with an average Fe yield per SNII $<y_{\rm Fe}>\sim 0.1$ M$_{\sun}$ (e.g.
Gibson, Loewenstein, \& Mushotzky 1997), the averaged metallicity
produced by all SNII is $<Z_{\rm Fe}>_{\rm SNII}\sim 0.053\nu$ solar.
To make the observed metallicity, a high SNII production
efficiency is needed: $\nu\sim 5-6$ (this value would be
reduced somewhat if SNIa contribute a significant fraction of iron).
Thus, it appear that both heating and cosmic metallicity may be produced
by a stellar population with $\nu\approx 5$.
The agreement between these two estimates of $\nu$ supports SNII as the
source of non gravitational heating. However, it should be noted that
such a large production of SNII may be inconsistent with the chemical
evolution of ISM in elliptical galaxies (Brighenti \& Mathews 1999b).

\section{Conclusions}

{\it External heating} models fit the data in the $L-T$ plot 
provided the preheating
is sufficiently strong: $\sim 1-4$ keV/particle, depending on $z_{\rm h}$.
However, successful models in the $L-T$ plane have entropies that exceed
observed values (a cautionary note: we are well aware that our models are 
approximate, and it's possible that less extreme preheating scenarios
may overcome the problems pointed out by the present work).
The competing models with {\it internal heating} by SNII fit
the whole set of X-ray observations better and more plausibly, but they
likely require a production of energy per unit of stellar mass larger
than that based on a Salpeter IMF. This may not be a severe demand
since the cosmic metallicity itself requires such a higher number of SNII 
per unit of stellar mass.

\end{document}